# Atomistic simulations of nematic phases formed by cyano-biphenyl dimers.

Alexandros G. Vanakaras[1] and Demetri J. Photinos

*Department of Materials Science, University of Patras, Patras, Greece*

**Abstract**

Molecular dynamics simulations of selected members of the cyano-biphenyl series of dimers (CBnCB) have been set up using atomistic detail interactions among intermolecular pairs of united atoms and allowing fully for the flexibility of the spacer chain. Detailed results are presented for the CB7CB dimers, showing an isotropic fluid phase and two nematic phases. The positional and orientational correlation functions extracted from the simulations are used to elucidate the structure of the low-temperature nematic phase. Polar molecular ordering is clearly identified along a direction undergoing helical twisting at right angles to the helical axis, with a constant pitch of about of 8nm. The local ordering of the various molecular segments is calculated and found to be in excellent agreement with experimental NMR measurements. Key findings of the simulation are shown to be correctly predicted by the theoretical model of the polar-twisted nematic ($N_{PT}$) phase [A.G. Vanakaras, D.J. Photinos, Soft Matter. 12 (2016) 2208–2220]. The complete failure of the usual twist bend model ($N_{TB}$) to account for these findings is demonstrated.

**1. Introduction**

Thermotropic nematics formed by achiral low molar mass mesogens constitute the structurally simplest liquid crystal (LC) phase and, to date, are the most widely used in electro-optic applications of LCs in the display technology. This nematic phase, hereafter denoted by $N_U$, shows no long-range order of the molecular positions but is characterized by uniaxial apolar ordering of the molecular orientations about a unique direction, represented by a unit vector **n**, the nematic phase "director". The uniaxiality of the phase implies full rotational symmetry (i.e. invariance of all the physical properties) about the director whilst the a-polarity of $N_U$ implies the physical indistinguishability of the directions **n** and –**n**. The lowest free energy state of the bulk $N_U$ phase corresponds to spatially uniform director field. Mild deformations of the director field, preserving the local symmetry of the phase and the degree of molecular ordering, give rise to elastic behavior of the bulk $N_U$; three types of such director-field deformation are possible: bend, splay and twist.[1]

[1] a.g.vanakaras@upatras.gr, photinos@upatras.gr

Despite theoretical predictions suggesting the possibility of various deviations from the $N_U$ phase, corresponding to biaxial nematic phases ($N_B$) [2], polar nematic phases, [3,4] uniaxial [5] ($N_{PU}$) or biaxial[6] ($N_{PB}$), spontaneous twist-bend ($N_{TB}$) nematics [7], etc, the $N_U$ phase remained until recently the only stable thermotropic nematic phase that could be produced experimentally. Thus, unlike the rich polymorphism found in the smectic phases or in the columnar phases of LCs [8,9], all known thermotropic nematics belonged to the $N_U$ phase and its chiral counterpart (N*). Chiral nematic phases were known to be produced only by chiral mesogens. Such phases usually show spontaneous twist of the director **n**, with helical pitch often in the optical wavelength regime. Close to the isotropic phase (I) transition temperature, the helical structures may show three-dimensional topological order, leading to the blue phases (BPs).

The above landscape of the nematic phase, comprising only liquid crystals of the $N_U$ phase, presently termed as "common nematics", and the chiral N* (also known as cholesteric), has changed significantly, and somehow surprisingly, as a result of novel experimental findings in the last 15 years: (i) The long predicted biaxial nematic phase appeared to have been finally identified experimentally in thermotropic bent-core nematogens[2,10]. However, it became evident soon thereafter, that, unlike the idealized orthorhombic monodomain biaxial nematic phase envisaged in the original theories, the experimentally identified phase could be described as consisting of cybotactic biaxial clusters, [11–16] most likely of monoclinic symmetry. The macroscopic biaxiality of such phase would not appear spontaneously but as an induced effect under the action of external fields of relatively low strength. (ii) Furthermore, as the molecular ordering within the cybotactic clusters formed by certain classes of bent-core mesogens could be polar, the collective alignment of such clusters gives rise to field-induced, macroscopically polar, nematic media capable of showing ferroelectric switching [6,17,18]. (iii) The continuing search for stable biaxial thermotropic nematics lead to a yet more surprising discovery[19]: a positionally disordered phase formed by symmetric achiral mesogenic dimers which, however, shows chiral ordering of the molecules in helical domains of opposite handedness[20–29]. The first observation of such phase, identifying it as a second, low temperature, nematic phase of odd-spacer CBnCB dimers, dates back to 1991 [30]. In a subsequent publication, the same phase was reported as a smectic[31] of unspecified structure. It was only in 2010 that this phase was conclusively reported as a positionally disordered and structurally chiral phase[20], for which the symbol Nx was tentatively introduced.

The discovery of the Nx phase initiated intensive, and still ongoing [32–37], investigations of its properties and thereby of its structure and distinction from other LC phases. It is now generally accepted that the orientational order in this phase is helically modulated with pitch on the order of 10 nm[23,38,39] and with the possibility of coexistence of both senses of helical twisting in different domains within an unbiased sample[20,40]. However, the molecular ordering and local symmetries of the phase are still open to debate [41–43] which ranges from adopting for the Nx phase the twist-bend nematic model ($N_{TB}$)[44,24,45], originally introduced by R.B. Meyer in 1974[7], to suggestions that the Nx should not be termed as a nematic phase[27,46]. The view we adopt in this paper, and subject it to testing with atomistic resolution molecular simulations, is that: (i) The local symmetry of the Nx phase differs from that of the $N_U$ phase, and therefore the $N_{TB}$ model, being based on the spontaneous spatial modulation of the nematic director **n,** is simply not applicable. (ii) The local molecular organization within the Nx phase is positionally disordered and orientationally strongly polar; the direction of polar ordering (denoted by the unit vector **m**) constitutes the only "phase director" (i.e. principal axis of all tensor properties) which is also an axis of twofold symmetry. (iii) In the undistorted state the polar director **m** is spatially modulated forming a constant-pitch helix and remaining perpendicular to the helix axis direction, denoted by the unit vector **$n_h$**.

Naturally, such phase, lacking a "nematic director", does not fall within the current classification of nematics (uniaxial or biaxial); accordingly the question is raised as to whether the present classification should be broadened or new classes of positionally disordered liquid crystals should be allowed for. Bypassing this issue for the moment, we will, for brevity, use for the above described model-phase the name "polar twisted nematic" (in short $N_{PT}$). This phase was shown to be thermodynamically stable in the context of a molecular theory[43] that has been formulated on the basis of a minimal molecular model embodying the essential features (bent shape, flexibility, statistical symmetry) of the odd-spacer CBnCB dimers. The local symmetry of the resulting thermodynamically stable phase has been successfully tested against NMR experimental spectra on rigid probe solutes[47] in the Nx phase of dimers.

In this paper we present atomistic detail simulations of the CBnCB dimers, which corroborate the predictions of the theoretical $N_{PT}$ model and also provide experimentally relevant information which could not be obtained from the theoretical model due to the minimal representation of the molecular structure employed therein[43]. Computer simulations of dimer molecules models at atomistic details have been reported in the past[48] but showed no LC phase with the characteristics of Nx. On the other hand, simulations of model bent-core molecules, either completely rigid [49,50] or allowing for internal rotations of the core segments and mimicking the flexibility of pendant end chains [51], have also been reported and show helically structured LC phases. Thus, to our knowledge, the work presented here reports for the first time atomistic simulations of mesogenic dimers that allow fully for the flexibility of the spacer and show clearly a thermodynamically stable phase with the characteristics of the Nx.

The remainder of the paper is structured as follows: The molecular organisation, the symmetry and the main physical properties of the phase described by the polar twisted model, $N_{PT}$, are outlined in section 2. The molecular simulation of the CBnCB dimers is presented in Section 3. Simulations were prepared for CBnCB series with spacer lengths in the experimentally relevant range n=7 to 11. In this work we do not attempt an exhaustive presentation of these simulations; rather we focus on the results obtained for the CB7CB dimers. Section 4 contains the results of the simulations and a discussion of their relevance to theoretical predictions, existing experimental findings on CB7CB and predictions for future experimental investigation. The conclusions of this study are presented in section 5.

## 2. Molecular organisation in the polar twisted nematic ($N_{PT}$) model.

The polar twisted model emerged from a molecular theory[43] of the positionally disordered fluid phases that can be formed by idealized molecules mimicking the key features and symmetries of the CBnCB mesogenic dimers. In particular, the directional interactions of the molecules originate from two rod-like mesogenic units that are separated by a non-interacting spacer and are allowed to assume different relative dispositions defining a statistically achiral set of opposite chirality conformations which share a common $C_2$ symmetry axis. The theory shows that three such phases can exist as thermodynamically stable: the high temperature phase is an isotropic fluid, the low temperature is the twisted polar director nematic and the intermediate temperature phase is the usual, uniaxial, nematic phase. The thermodynamic stability of the latter phase is lost in certain regions of the molecular parameterisation, giving rise to direct phase transitions from the isotropic fluid to the polar twisted nematic (I-$N_{PT}$ transitions).

Unlike the uniaxial nematic, the polar twisted nematic phase has no axis of full rotational symmetry but only a local $C_2$ axis, the "polar director", denoted by the unit vector **m**, with **m** and −**m** corresponding to physically distinguishable states. The minimum free energy configuration of the **m**-director field is a helix of uniform pitch with **m** remaining perpendicular (Figure 1) to the constant helix axis denoted by the direction $\mathbf{n_h}$. Left- and right-handed helical twisting of **m** are equally probable in an unbiased sample due to the statistical achirality of the constituent dimer molecules. The helical twisting of **m** is rather tight: the pitch is on the order of a few molecular lengths and shows marginal variation with temperature after reaching rapidly its full value within a narrow interval below the phase transition temperature. The helical environment exerts a chiral bias to the conformations of the dimer molecules but the deviation from the strict statistical achirality (quantified by an "induced molecular chiral asymmetry parameter") is rather small. In contrast, the polar ordering of the molecular $C_2$ axis along the director **m** is strong and defines the primary polar order parameter $p$ of the phase[43].

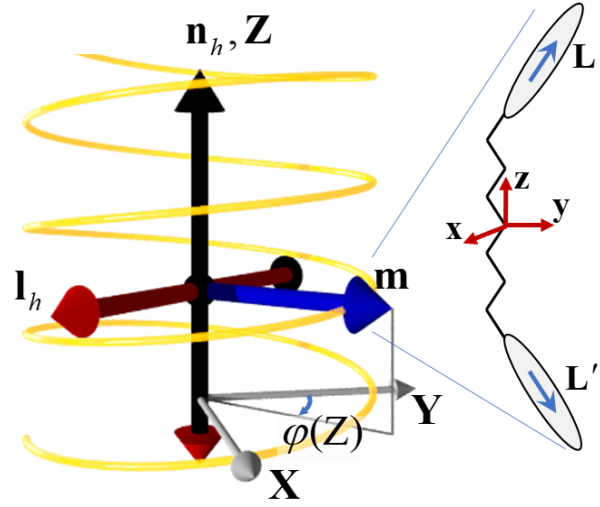

**Figure 1.** Schematic representation of the local molecular organisation and the modulation of ordering in the polar twisted nematic ($N_{PT}$) model-phase.

The second rank ordering tensors of the various molecular segments ($s$) have **m** as a common principal axis. The directions of the other two principal axes are on the plane perpendicular to **m** and from angles $\theta_s$ and $\theta_s+\pi/2$ with the helix axis $\mathbf{n}_h$. The angle $\theta_s$, often referred to as the "tilt" angle associated with the molecular segment $s$, is clearly not a uniquely defined property of the phase as it may change from segment to segment of the same molecule.

The twisting of the **m** director about $\mathbf{n}_h$ implies a respective helical twisting of the other principal axes associated with the molecular segmental ordering tensors. However, while the twisting of **m** is at right angles to the helix axis $\mathbf{n}_h$, the other principal axes undergo heliconical twisting forming angles $\theta_s$ and $\theta_s+\pi/2$ with $\mathbf{n}_h$. This characteristic bears a superficial resemblance to the twist bend model ($N_{TB}$) of the nematic phase[7]. Such resemblance, however, should not cause any confusion as the two models have very clear and fundamental differences: The $N_{TB}$ model[7,44] is obtained within the Frank-Oseen elasticity of uniaxial nematics[1] wherein, under certain conditions for the elastic constants, a modulation of the nematic director **n** (i.e. of the local axis of full rotational symmetry of the phase) can emerge which shows simultaneously constant twist and constant bend deformations[7,44,52,53]. The modulation in this case is heliconical, with the twisting-bending nematic director **n** forming a constant "tilt" angle with the helix axis **h.** Obviously, such tilt angle characterizes the $N_{TB}$ phase as a whole, in other words, it is the same for all molecular segments. This is quite distinct from the helical modulation characterizing the polar twisted phase, $N_{PT}$, wherein (i) no axis of full rotation symmetry (i.e. nematic director) exists, (ii) the molecular ordering is strongly polar about director **m,** (iii) the polar director **m** is a local $C_2$ axis of the phase, (iv) **m** is the twisting entity (v) the tight twisting of **m** is a result of molecular packing at equilibrium and not of any elastic deformation that preserves the local ordering within the phase. Naturally, in the $N_{TB}$ model, the bend vector $\mathbf{n} \times (\nabla \times \mathbf{n})$ defines a polar direction at all points of the phase. Such polarity, however, characterizes the <u>deformation</u> of **n,** <u>not the molecular ordering,</u> since the latter remains apolar and fully symmetric about **n** at all points of the $N_{TB}$ phase, just as it does in the undeformed uniaxial nematic ($N_U$) phase.

## 3. Molecular simulation of the CB7CB dimer.

### *3.1 Molecular model and force field.*

The mesogenic units of the CBnCB dimers are represented as rigid linear arrays of seven Lenard-Jones (LJ) beads. The potential between two LJ beads, 1 and 2, separated by a distance $r$ is given according to the standard expression

$$u(r) = 4\varepsilon \left( \left(\frac{\sigma}{r}\right)^{12} - \left(\frac{\sigma}{r}\right)^{12} \right) . \quad (1)$$

Based on the effective dimensions of the molecular segments forming the CB mesogenic units, four LJ sites, separated by a constant distance of $2.3\,\text{Å}$, are used to represent the two phenyl rings. The CN group is represented by two LJ identical sites separated by a constant distance of $1.3\,\text{Å}$. The $CH_2$ units at the $\alpha, \omega$ positions of the dimer are also part of the rigid mesogenic unit. Both the $CH_2$ site and the C atom of the CN group are separated by a constant length of $2.2\,\text{Å}$ from the corresponding phenyl LJ sites. In this representation the distance between the $\alpha, \omega$ $CH_2$ groups and the terminal N-atoms of the respective mesogenic cores is $12.6\,\text{Å}$; this is slightly longer than the length of the mesogenic unit found according to standard molecular mechanics calculations ($\sim 11.5\,\text{Å}$). The mass of each united atom is set equal to the molecular weight of the corresponding group, i.e. $m_C = 12$, $m_N = m_{CH_2} = 14$ and $m_{ph} = 38$. The chosen values for the diameters of the various sites are $\sigma_{CH_2} = 3.90\,\text{Å}$, $\sigma_C = \sigma_N = 0.90\sigma_{CH_2}$ and $\sigma_{ph} = 1.30\sigma_{CH_2}$. The LJ parameters $\varepsilon$ are $\varepsilon_C = \varepsilon_N = \varepsilon_{CH_2} = 75$ K and $\varepsilon_{ph} = 1.25\varepsilon_{CH_2}$. The standard geometric mixing rule is applied for the parameters of the interaction potential between dissimilar sites and a cut-off $2.5\sigma$ is used. The bond potential between successive $CH_2$ sites of the spacer is of the form $u_b(l) = k_b(l - l_0)^2$ with $k_b = 60000\,\text{K/Å}^2$ and $l_0 = 1.53\,\text{Å}$; the bond angle contribution is of the form $u_\theta(\theta) = k_\theta(\theta - \theta_0)^2$ with $k_\theta = 20000\,\text{K/rad}^2$ and $\theta_0 = 114°$. For the dihedral angles we have used the Ryckaert & Bellemans torsional potential. [54] This torsional potential, and similar parameterization of the spacer, has also been used for the simulation of CB8CB [48].

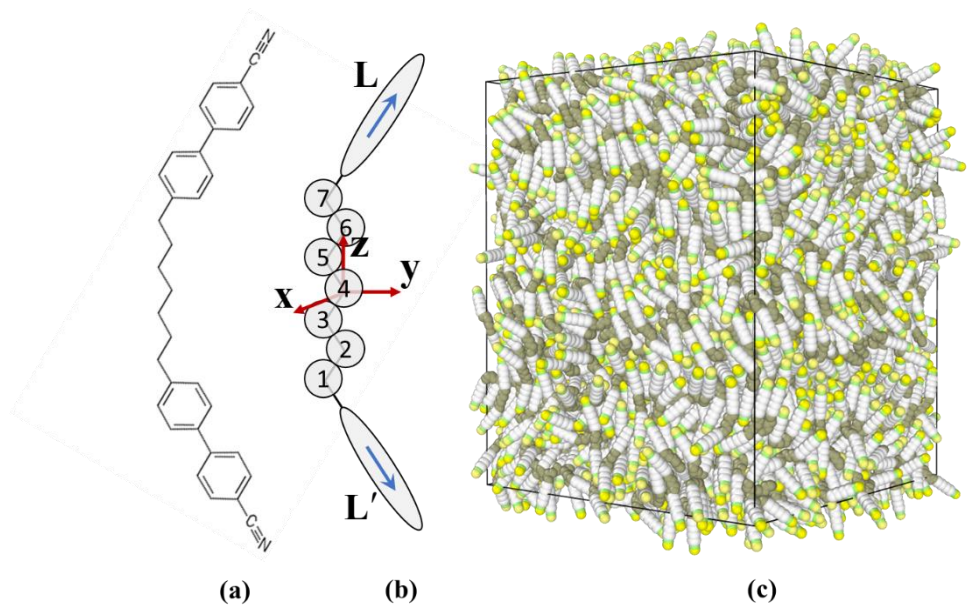

**Figure 2.** (a) Chemical structure of the CB7CB dimer molecule (b) Coarse-grain representation used in the simulations. The mesogenic units are assumed rigid and uniaxial, with their directions denoted by the unit vectors **L** and **L**′. The shown molecular frame of axes **x**, **y**, **z** is rigidly attached to the central CCC group of the spacer. (c) Typical snapshot of a simulated system of $N$=3072 dimers in its (low temperature) nematic state.

*3.2 Simulation details.*

The MD simulations were performed with the open source software LAMMPS[55]. We simulated systems of various sizes, ranging from $N = 2000$ to 12288 CB7CB dimers, both in the *NVT* in the *NpT* ensembles assuming periodic boundary conditions. Starting from a well equilibrated configuration in the isotropic liquid state of the system at a relatively high temperature, we have cooled the system slowly. During this first cooling cycle, the systems was barostated with the Berendsen method and the degree of orientational order of the system was monitored. Long runs (corresponding to 20-50 ns) were performed at temperatures where the system starts to develop orientational order. Taking into account that, on cooling from the isotropic state, the appearance of long range orientational order is accompanied by the tendency of the systems to develop structural chirality, the size of the simulation box was judiciously chosen so as to yield systems able to accommodate states with two or more full helical pitches. To avoid the appearance of helical states with artifact pitch, generated by the periodic boundary conditions, we chose the length of one of the sides of the simulation box to be a multiple of the expected helical pitch of CB7CB (~8nm). In this way we were able to generate well equilibrated, defect-free states with the helical axis of the ordered phase parallel to one of the axes of the simulation box, assigned as the *Z*-axis. These states were then used as initial states to generate heating and/or cooling sequences. The subsequent changes of box dimensions, as allowed in the *NpT* ensemble, enable the system to reach the equilibrium helical pitch.

We note here that the lateral (*X, Y*) side lengths of the simulation box should be at least twice the distance beyond which the respective (side-by-side) positional correlations between the dimers become negligible. This correlation length, even for the high temperature twisted nematic states, is in the range 30-40Å, indicating that the length of the *X,Y* box sides should exceed 80Å. For smaller lateral dimensions, artificial self-enhancement of the positional and orientational correlations is observed, leading, in addition to spurious correlations and other severe size effects, to a tendency of the helical axis to deviate from the *Z*-axis of the simulation box. Such behavior is found to disappear for simulated system sizes of $N > 2000$ dimers.

The simulated CB7CB dimers exhibit three distinct phases appearing in the phase diagrams of figure 3: the isotropic (I) and two nematic phases. Both nematic phases are macroscopically (i.e. for samples of the size of the entire simulation box) uniaxial and apolar but they are clearly distinguished since the low temperature nematic phase exhibits structural chirality and spatial modulation of the orientational order. In this study we shall focus mainly on the analysis of the structure of the low temperature nematic phase of the CB7CB dimer.

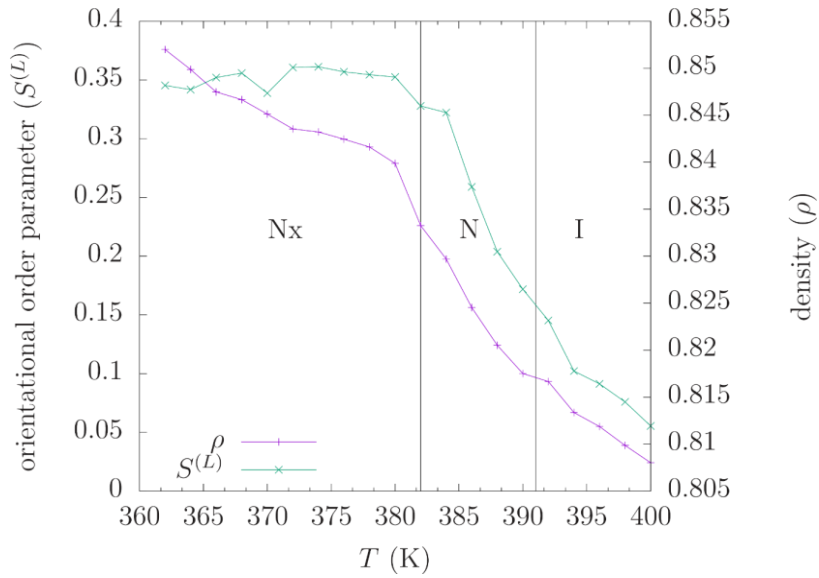

**Figure 3**. Calculated temperature dependence of the orientational order parameter ($S^{(L)}$) of the mesogenic units and of the density (*ρ*) for a simulated system of *N*=3072 CB7CB dimers barostated at pressure $p = 15$ atm. The vertical lines indicate the range of stability of the successive phases obtained.

### *3.3. Physical quantities extracted from the simulations.*

The ordering tensor of any unit vector associated with a given molecular segment, *s* (mesogenic unit, C-C or C-H bond, inter-nuclear vector, etc) is defined as

$$Q_{ab}^{(s)} = \frac{1}{2}\left\langle \sum_i \left(3 e_{i;a}^{(s)} e_{i;b}^{(s)} - \delta_{ab}\right)\right\rangle , \qquad (2)$$

with the *i*-summation extending over all the molecules in the ensemble, $e_{i;a}^{(s)}$ denoting the component of the unit vector of the molecular segment *s* along the $a = X, Y, Z$ axis of the simulation box. The eigenvectors of the $Q_{ab}^{(s)}$ tensor give the principal axis frame (*PAF*) associated with the orientational ordering of the *s*-segment, and the corresponding eigenvalues quantify the degree of ordering in the *PAF*. In a locally uniaxial medium the *PAF* of the ordering tensor of any segment, or, equivalently, of any second rank physical property, share a common axis of ordering, the local uniaxial "nematic director". In that case, the common principal axis of the ordering tensors associated with the mesogenic units and with any other molecular segment, together with the equality of the eignevalues associated with the other two principal axes, confirm that the system is locally uniaxial. The polarity of the phase is monitored through the segmental vector ensemble averages ("polar order parameters")

$$p_a^{(s)} = \left\langle \sum_i e_{i;a}^{(s)} \right\rangle . \qquad (3)$$

To quantify the positional correlations, we have calculated the dependence of the usual radial pair correlation function between the centers of mass of the molecular segments *s* on the projection of the inter-segment distance parallel ($r_\parallel$) and perpendicular ($r_\perp$) to the unique macroscopic symmetry axis of the phase, $\mathbf{n}_h$. These functions are defined as

$$g_\parallel^{(s)}(r_\parallel) = \left\langle \delta(r_\parallel - \hat{\mathbf{n}}_h \cdot \mathbf{r}_{ij}^{(s)}) \right\rangle_{ij} \text{ and } g_\perp^{(s)}(r_\perp) = \left\langle \frac{1}{r_\perp} \delta(r_\perp - \sqrt{r_{ij}^{(s)} - \left(\hat{\mathbf{n}}_h \cdot \mathbf{r}_{ij}^{(s)}\right)^2}) \right\rangle_{ij}, \qquad (4)$$

where the angular brackets indicate time averaging over all intermolecular pairs *ij*.

Representative plots of the calculated $g_\perp^{(s)}(r_\perp)$ for the central carbon of the spacer as well as for the center of mass of the mesogenic end-segments of the simulated CB7CB dimer at the temperature of $T = 380\text{ K}$ are given in figure 4. A small degree of side-by-side positional correlations is observed, diminishing above 30Å. The respective $g_\parallel^{(s)}(r_\parallel)$ show marginal variation about the value of 1, throughout the *Z*-range of the simulation box. Accordingly, it is concluded that the system is positionally disordered at that temperature.

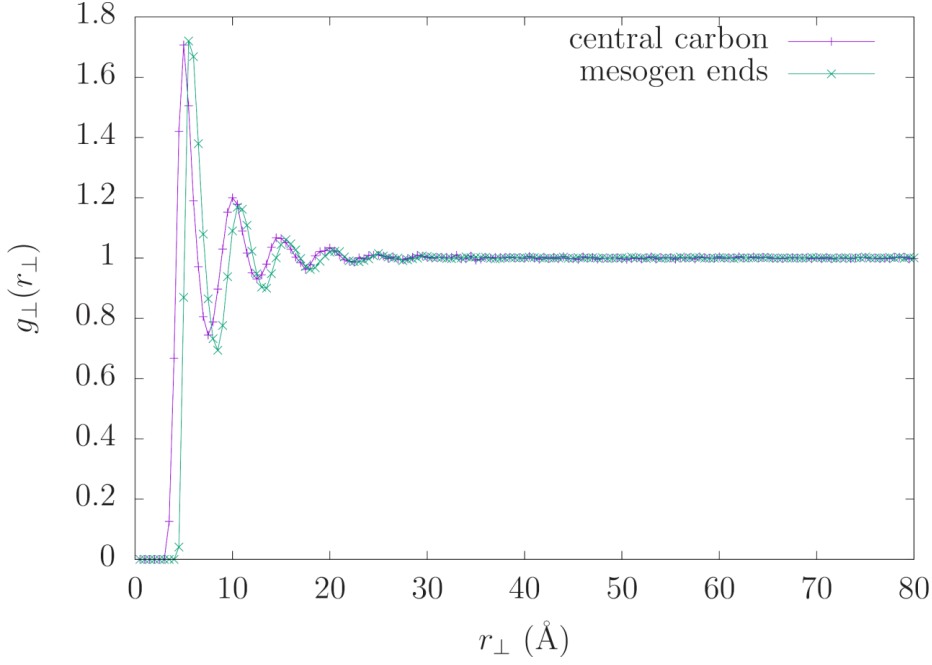

**Figure 4.** Positional correlations between the central carbons of the dimer (+) and the center of mass of the mesogens (x), as functions of the distance perpendicular to the helix axis ($\mathbf{n}_h$) of the low temperature nematic phase of the simulated CB7CB dimers at $T = 370$ K for a system of $N$=12288 dimer molecules.

Clearly, the absence of any long-range positional correlations, either along or perpendicular to the unique macroscopic axis of the phase, are typical of common uniaxial nematic phases but do not allow the extraction of further inferences regarding the local structure of the phase. Such inferences can be reached through the distance dependence of the orientational correlations of low tensor rank. Analogously to eq(4), the $m^{th}$-rank orientational correlations of the various intermolecular segment pairs as functions of the projections of their distance along and perpendicular to the axis $\mathbf{n}_h$ of the phase are calculated, respectively, as

$$g_{m,\parallel}^{(s)}(r_\parallel) = \left\langle P_m\left(\mathbf{e}_i^{(s)}(\mathbf{r}_i)\cdot\mathbf{e}_j^{(s)}(\mathbf{r}_j)\right)\delta(r_\parallel - \hat{\mathbf{n}}_h\cdot\mathbf{r}_{ij}^{(s)})\right\rangle_{ij} / g_\parallel^{(s)}(r_\parallel) \qquad (5)$$

and

$$g_{m,\perp}^{(s)}(r_\perp) = \left\langle P_m(\mathbf{e}_i^{(s)}\cdot\mathbf{e}_j^{(s)})\delta(r_\perp - \sqrt{r_{ij}^{(s)} - (\hat{\mathbf{n}}_h\cdot\mathbf{r}_{ij}^{(s)})^2})\right\rangle_{ij} / g_\perp^{(s)}(r_\perp) \qquad (6)$$

Here $P_m(x)$ are the Legendre polynomials of rank $m$; $\mathbf{e}_i^{(s)}(\mathbf{r}_i)$ is a unit vector belonging to molecule $i$ and rigidly attached to molecular segment $s$ located at $\mathbf{r}_i^{(s)}$ and $\mathbf{r}_{ij}^{(s)}$ is the vector connecting the $s$-segments of molecules $i$ and $j$. Results on the orientational correlations of ranks $m = 1$ and $2$ are presented and analyzed in the next section.

Lastly, we have monitored several parameters connected with the average shape of the dimer [56] during the simulations. Such temperature dependent shape parameters are rotationally invariant and provide specific insights on the conformational changes of the dimer across the three fluid phases of the simulation. To quantify how the twisted structure disturbs the statistical achirality of the dimer we have calculated the molecular shape parameter

$$c = \left\langle \left( \hat{\mathbf{L}} \times \hat{\mathbf{L}}' \right) \cdot \hat{\mathbf{d}}_{LL'} \right\rangle , \qquad (7)$$

with $\hat{\mathbf{L}}, \hat{\mathbf{L}}'$ denoting the orientations of the mesogenic ends (rod-like molecular units) and $\hat{\mathbf{d}}_{LL'}$ the unit vector along the direction connecting their geometrical centers. This parameter (also referred to as "induced chiral asymmetry" parameter[43]) provides a measure of the probability imbalance between the enantiomeric sets of the molecular conformations, induced by the chirally modulated directional environment of the flexible dimer molecule.

## 4. Results and discussion.

The results presented in this section are for a well equilibrated system composed of $N=12288$ CB7CB dimers, simulated for more than 50 ns, at constant volume ($\rho = 0.845 \text{ g/cm}^3$) and temperature $T = 380 \text{ K}$. According to the discussion in section 3.2, the relatively large number of molecules contained in the simulation box ensures that size effects on the calculated results are minimal.

To analyze the local structure through the orientational correlations we have calculated pair correlations associated with the axes $\mathbf{x}, \mathbf{y}, \mathbf{z}$ of a molecular frame attached to the central carbon of the dimer spacer. The $\mathbf{x}$ unit vector is normal to the plane of the CCC group and $\mathbf{y}$ is along the bisector of the central CCC group angle (see figure 2). The $\mathbf{y}$ molecular direction is the "statistical" $C_2$ symmetry axis of the isolated dimer, since a rotation by $\pi$ about this axis generates a statistically equivalent isomeric molecular state. The $\mathbf{z}$ axis is along the direction connecting the two outer carbons. We have calculated also the orientational correlations associated with the orientations $\mathbf{L}$ of the mesogenic units (Figure 2).

The results for the calculated first rank (polar) correlations according to eq (6) are presented in Figure 5(a), where it is apparent that $g_{1,\perp}^{(s)}(r_\perp)$ does not decay to zero for the molecular vectors $\mathbf{L}$ and $\mathbf{y}$. This demonstrates clearly the existence of long-range polar order along a (polar) "director" $\mathbf{m}$ that is perpendicular to $Z$. The long-range polar correlations are exhibited only by molecular vectors having a non-vanishing component along the molecular $C_2$ axis. For molecular vectors perpendicular to the $C_2$ axis (i.e. the $\mathbf{x}, \mathbf{z}$ molecular vectors) the polar correlations are absent at all distances.

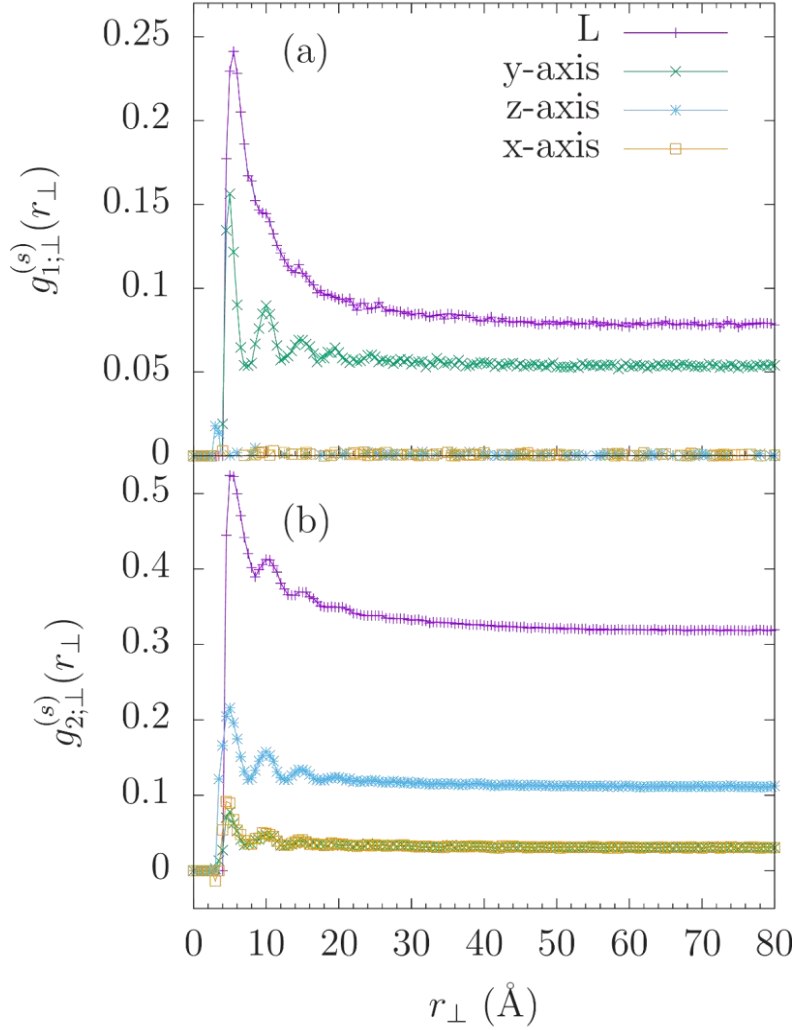

**Figure 5.** Representative polar (a) and quadrupolar (b) orientational correlations perpendicular to $\mathbf{n}_h$, calculated according to eq(6).

On the other hand, the calculated quadrupolar orientational correlations (corresponding to $m=2$ in eq (6)), shown in figure 5(b), are long ranged for any molecular segment. The plots in this figure indicate that the mesogenic units are the most ordered molecular segments while the quadrupolar ordering of the molecular $\mathbf{y}$-axis is seen to be higher than the corresponding ordering of the $\mathbf{x}, \mathbf{z}$ axes.

Collecting the inferences from the results plotted in figures (4) and (5) for the simulated low temperature fluid phase we conclude that the phase is (i) positionally disordered (ii) macroscopically uniaxial and overall apolar and (iii) clearly characterized by long range polar correlations perpendicular to its macroscopic symmetry axis. These inferences then pose the question of identifying the mechanism by which a transversely polar system is rendered macroscopically uniaxial and apolar. This mechanism is revealed by studying the polar and quadrupolar orientational correlations, $g^{(s)}_{m,\parallel}(Z)$, along the $\mathbf{n}_h$ phase axis.

In Fig 6(a) we present the polar correlations along the macroscopic Z-axis for the $\mathbf{x}, \mathbf{y}, \mathbf{z}, \mathbf{L}$ vectors, calculated from the simulations according to eq(5). As in the case of the radial correlations perpendicular to Z, the $\mathbf{y}$ and $\mathbf{L}$ molecular directions exhibit non-vanishing polar correlations, the strength of which follows a well-defined periodic oscillation with a periodicity of $\sim 82\text{Å}$. This suggests that the polar director change its direction with Z ($\mathbf{m}=\mathbf{m}(Z)$), twisting uniformly at constant pitch about the $\mathbf{n}_h$ axis and remaining perpendicular to the latter. The twisted structure of the phase is also confirmed by the quadrupolar orientational correlations along Z presented in Fig. 6(b,c).

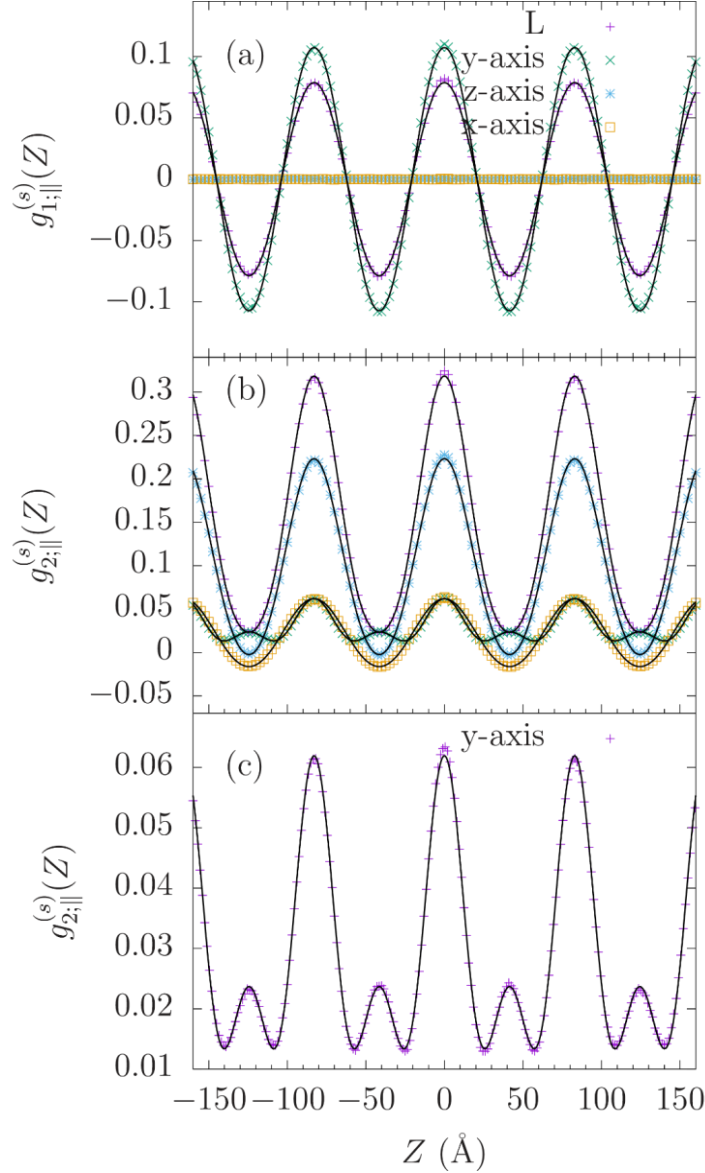

**Figure 6.** Polar (a) and quadrupolar (b,c) orientational correlations along $\mathbf{n}_h$ for the mesogenic units and the central carbon molecular axes calculated according to eq(6) from the simulations of CB7CB dimers. The simulation values are shown by the indicated symbols. The solid lines are obtained by fitting the respective simulation points according to the director modulations introduced in the $N_{PT}$ model. Such fits yield the optimal values of the pitch of the helix, the tilt and the local orientational order parameters exhibited by the different segments.

These results confirm the helical modulation of the phase. However, their rationalization in terms of the local molecular ordering necessitates an analysis of the ordering within thin slabs centered about a fixed value of the Z coordinate of the simulated system. The net Z dependence of the correlations $g_{m,\|}^{(s)}(Z)$ among the molecules in given slabs is obtained by integrating over the lateral coordinates $X,Y$, of the simulation box. Such integration then extends over macroscopic regions, wherein the orientations of the segmental pair are uncorrelated, except for a small area whose size is determined by the respective correlation length. Accordingly, introducing explicitly in eq(5) the local axes frames of the slabs to which a pair of molecules belongs, the first and second rank correlations can be but in the form

$$\tilde{g}_{1,\|}^{(s)}(Z) = \left\langle \overline{e}_a^{(s)}(Z')\overline{e}_{a'}^{(s)}(Z'')\big(\mathbf{a}(Z')\cdot\mathbf{a}'(Z'')\big)\delta(Z-(Z'-Z''))\right\rangle$$

$$\tilde{g}_{2,\|}^{(s)}(Z) = \frac{3}{2}\left\langle \overline{e}_{ab}^{(s)}(Z')\overline{e}_{a'b'}^{(s)}(Z'')\big(\mathbf{a}(Z')\cdot\mathbf{a}'(Z'')\big)\big(\mathbf{b}(Z')\cdot\mathbf{b}'(Z'')\big)\delta(Z-(Z'-Z''))\right\rangle - \frac{1}{2}, \quad (8)$$

with $\mathbf{a}(Z),\mathbf{b}(Z)$ denoting the local phase frames of phase axes at $Z$ and the "slab averages"

$$\overline{e}_a^{(s)}(Z) = \sum_{i=1}^{N} e_{i;a}^{(s)}\delta(Z-Z_i) / \sum_{i=1}^{N}\delta(Z-Z_i),\quad \text{and}$$

$$\overline{e}_{ab}^{(s)}(Z) = \sum_{i=1}^{N}\big(e_{i;a}^{(s)}e_{i;b}^{(s)}\big)\delta(Z-Z_i) / \sum_{i=1}^{N}\delta(Z-Z_i),\quad \text{representing the polar and quadrupolar}$$

components of the segmental ordering tensors within a slab centered at $Z$. Therefore, the expressions in eqs (8) provide the orientational correlations between the local "slab-average" ordering within parallel slabs separated by a distance Z along the helix axis.

Attempting to interpret the simulation findings in figures (5) and (6) according to the twist-bent nematic ($N_{TB}$) model, we note that, within that model, each slab is described by a single director $\hat{\mathbf{n}}$ which rotates at constant pitch, remaining constantly tilted by an angle $t$ with respect to the helical axis of the phase ($\mathbf{n}_h$): $\hat{\mathbf{n}}(Z) = (\cos kZ \sin t, \sin kZ \sin t, \cos t)$. Clearly the assumed symmetries of this model lead to $\overline{e}_a = 0$, for all phase axes and molecular segments, consequently implying $g_{1,\|}^{(s)}(Z) = 0$ i.e. failing to account for any polar ordering within the slabs. Furthermore, the "tilt" angle $t$ characterizes the entire phase and not individual molecular segments.

On the other hand, within the polar twisted nematic ($N_{PT}$) model, the only "director" is the local polar axis of each slab. This axis follows the helical modulation $\hat{\mathbf{m}}(Z) = (\cos kZ, \sin kZ, 0)$, remaining perpendicular to the helical axis. In this model we have, according to eq.(8), $g_{1,\|}^{(s)}(Z) = p_{(s)}^2\big(\mathbf{m}(0)\cdot\mathbf{m}(Z)\big) = p_{(s)}^2 \cos kZ$ with $p_{(s)}$ the polar order parameter of segment $s$ along the $\mathbf{m}$ axis. In the graph of Fig.6(a) we present the simulated values of $g_{1,\|}^{(s)}(Z)$ together with their fitting according to the $N_{PT}$ model. The essentially exact fitting of the simulated values shown in this figure is obtained for the N$_{PT}$ model parameter $p_{(L)} = 0.27$, $p_{(y)} = 0.32$ and $d \equiv k/2\pi = 82.8$ Å for both, the mesogenic segment *L* and the central *y*-axis of the spacer (molecular $C_2$ axis).

Introducing in the second of eqs (8) the N$_{PT}$ model order parameters $S_h^{(s)} = (3\overline{e}_{n_h n_h} - 1)/2$, $\Delta_h^{(s)} = 2(\overline{e}_{mm} - \overline{e}_{l_h l_h})/3$ and $\delta_h^{(s)} = 2\overline{e}_{n_h l_h}/3$, with $\mathbf{l}_h(Z)$ denoting the phase axis perpendicular to $\mathbf{n}_h$ and $\mathbf{m}(Z)$, we get

$$g_{2,\parallel}^{(s)}(Z) = \left( (S_h^{(s)})^2 + \frac{3}{4}(\Delta_h^{(s)})^2 \cos 2kZ + 3(\delta_h^{(s)})^2 \cos kZ \right) \quad (9)$$

This equation is used to fit the calculated orientational correlations through eq. (8). The fitting parameters are, $S_h^{(s)}$, $\Delta_h^{(s)}$, $\delta_h^{(s)}$ and the periodicity $d \equiv k/2\pi$. The solid lines in in fig. 6(b),(c) are the fitting results, witch demonstrate essentially exact agreement of the N$_{PT}$ model predictions with the simulation results.

Noting that the director **m**, being a symmetry axis, is necessarily one of the three principal axes, the *PAF* of the ordering tensor of any molecular segment *s* is obtained by a rotation of the $\mathbf{n}_h$ and $\mathbf{l}_h$ axes by an angle $\theta^{(s)}$ ("segmental tilt"), with $\tan 2\theta^{(s)} = 2\delta_h^{(s)}/(S_h^{(s)} + \Delta_h^{(s)}/2)$ [43]. The best fitting parameters and the resulting $\theta^{(s)}$ for the various segments are summarized in table I.

**Table I** Best fitting parameters of eq (9) for the various molecular segments.

| Molecular segment axis (*s*) | $S_h^{(s)}$ | $\Delta_h^{(s)}$ | $\delta_h^{(s)}$ | $d$ | $\theta^{(s)}$ |
|---|---|---|---|---|---|
| **x** | 0.148 | 0.035 | 0.114 | 82.93 | 27.0° |
| **y** | 0.173 | 0.131 | 0.0799 | 82.92 | 16.9° |
| **z** | 0.320 | 0.101 | 0.194 | 82.94 | 23.1° |
| **L** | 0.390 | 0.158 | 0.222 | 82.95 | 21.7° |

Clearly, the segmental molecular ordering derived from the simulations implies distinctly different values for the rotation angles $\theta^{(s)}$ ("tilt") that diagonalize the ordering matrix of different molecular segments and/or axes thereof. This is in complete accord with the predictions of the N$_{PT}$ model and in sharp contrast with the unique "tilt" value implied by the N$_{TB}$ model for all the molecular segments.

Turning now to the influence of the chirally modulated orientational ordering on the conformational statistics of the CB7CB molecule we have evaluated from the simulations the chiral shape parameter $c$ of eq (7). In Figure 7 we present the probability distribution of $c$, namely $f(c) = \left\langle \delta\left(c - \left(\hat{\mathbf{L}}_i \times \hat{\mathbf{L}}'_i\right) \cdot \hat{\mathbf{d}}_{i;LL'}\right)\right\rangle_i$.

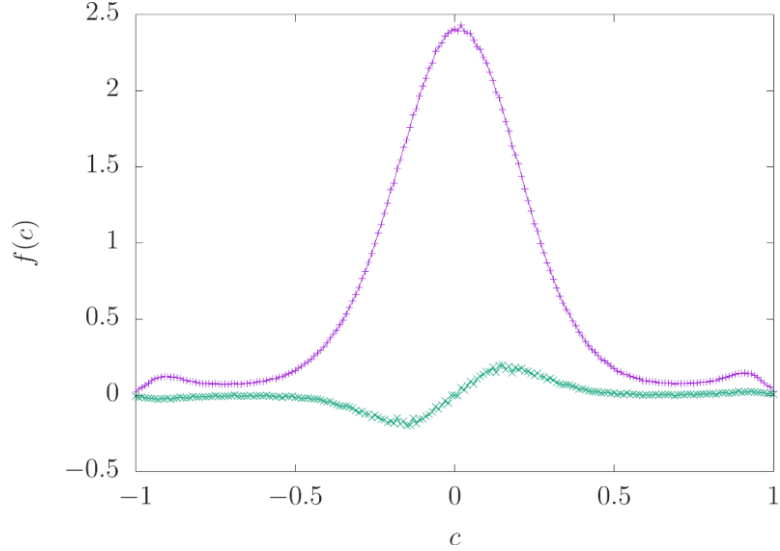

**Figure 7**. Probability distribution of the shape parameter $c$ which quantifies the induced chiral asymmetry on the conformations of the CB7CB dimer molecule in the chirally modulated phase. The lower curve shows the difference between the statistical weights of conformations with opposite handedness (i.e. the asymmetry of the top curve with respect to inversion of the sign of $c$).

Apparently there is a statistical imbalance between conformations of opposite handedness, but such imbalance is rather limited, in complete accordance with the predictions of the $N_{PT}$ model

An experimental signature of the Nx phase is the doubling of the NMR spectral lines associated with prochiral molecular sites of the dimer molecule or of probe solutes dissolved therein[25,47,57–59]. The phenomenon is known as enantiotopic discrimination and is commonly detected in chiral media. To test our simulation results against available experimental NMR measurements[24,60] on the Nx phase we have calculated the orientational order associated with the directions of the C-H inter-nuclear vectors. Assuming that the phase aligns with its macroscopic axis $\mathbf{n}_h$ parallel to the magnetic field of the NMR spectrometer ($\hat{\mathbf{n}}_h \parallel \mathbf{B}$) we have calculated segmental order parameters related to scaled spectral quantities[47,59] as follows:

(1) The ensemble averages $\tilde{v}_s = \left\langle P_2(\hat{\mathbf{e}}_{(CH)_s} \cdot \mathbf{n}_h)\right\rangle$ with $\hat{\mathbf{e}}_{(CH)_s}$ denoting the orientation of the C-H bond of the $s$-th methylene group of the spacer. These are proportional to the quadrupolar splittings $[\Delta v]_s$ of the C-D bonds of deuterated dimers, namely $\tilde{v}_s = [\Delta v]_s / (2v_{q;CD}/3)$, with $v_{q;CD}$ representing the quadrupolar coupling constant of the respective molecular segment; typically $v_{q;CD} = 165\,\text{kHz}$ for aliphatic C-D bonds.

(2) The ensemble averages $\tilde{D}_{s,s'} = \left\langle \dfrac{1}{r^3_{C_s-H_{s'}}} P_2(\hat{\mathbf{r}}_{C_s-H_{s'}} \cdot \hat{\mathbf{n}}_h) \right\rangle$, yielding the residual dipolar couplings of the inter-nuclear pairs formed by the $s-th$ carbon and the $s'$-th hydrogen; these parameters are sensitive both to the ordering within the phase and to the conformational statistics of the dimer. Their precise connection to the measured residual dipolar couplings $D_{s,s'}$ for the $^{13}C-^1H$ pairs is $\tilde{D}_{s,s'} = -D_{s,s'}/(3K_{CH}/2)$, with $K_{CH}$ denoting the dipolar coupling constant of the C-H nuclear pair ($K_{C-H} \approx 30\,\text{kHzÅ}^3$).

For the calculations, we have taken the plane formed by the HCH (or the DCD) group to be perpendicular to the plane of the corresponding CCC bond, the HCH angle to be 109.5° and the length of the CH bonds to be 1.1Å. The results for the residual dipolar couplings extracted from our simulations of the CB7CB dimers are summarized in Table II and are compared with available [60] experimental results (shown in parentheses).

**Table II**. Calculated ratios $\tilde{D}_{C_s,H_{s'}} / \tilde{D}_{C_4,H_4}$ for the C-H pairs of the spacer at the simulated temperature, with the simulated value for the central ($s=4$) C-H bond found at $\tilde{D}_{C_4,H_4} = -0.117$. The numbering of hydrogens indicates the carbon to which they are bonded according to figure 2. The unprimed hydrogens are on the same side of the spacer in its all-trans conformation. Their conjugates (primed) are in the opposite side. Shown in parentheses are the available experimental values $D_{C_s,H_{s'}} / D_{C_4,H_4}$ ( $D_{C_4,H_4} = 3247\,\text{Hz}$ ) obtained from ref [60] according to the assignment suggested therein. Experimental entries which have not been originally assigned in ref [60] are listed in parentheses in the last row.

|     | C1          | C2          | C3          | C4          |
|-----|-------------|-------------|-------------|-------------|
| H1  | 1.19 (1.13) | 0.15        | -0.04       | -0.03       |
| H1' | 1.55 (1.50) | 0.03        | -0.11       | -0.05       |
| H2  | 0.09        | 0.90 (1.16) | -0.12       | -0.13       |
| H2' | 0.01        | 1.07 (1.28) | 0.00        | -0.08       |
| H3  | -0.13       | -0.11       | 1.10 (0.89) | 0.08        |
| H3' | -0.07       | 0.03        | 1.21 (1.11) | -0.05       |
| H4  | -0.04       | -0.07       | 0.06        | 1.00 (1.00) |
| H4' | -0.04       | -0.12       | -0.06       | 1.00 (1.00) |
| H?  | **(0.13)**  | **(0.12)**  | **(0.13)**  | **(0.10)**  |

It should be noted that, although the simulation and measurement temperatures differ, the temperature variation of the residual dipolar couplings is small in the Nx phase and its influence is further reduced on the rations of couplings reported in the above table. It is apparent from the table that a better assignment, suggested by the simulation results, would be obtained by exchanging the experimental entries for the pairs C2-H2 and C2-H2´ with those of the C3-H3 and C3-H3´ pairs, respectively. Also the simulation results suggest possible assignments for the unassigned experimental entries in the last row. With such optimization of the assignments, it is evident that the agreement between the simulated and experimentally measured residual couplings is excellent.

The calculated quadrupolar NMR splittings are summarized in table III. There are no experimental results for the NMR spectra of dimers with fully deuterated spacers. The available experimental results[24,41] are for spacers deuteriated at their $\alpha,\omega$ positions. The experimental quadrupolar splittings for these positions (the $C_1D, C_1D'$ bonds in the numbering of Table III) are respectively $\sim 40$ and $\sim 55\,\text{kHz}$ at a temperature 37K below the N-I phase transition of CB7CB[24]. Our results demonstrate clearly the enantiotopic discrimination and are, for the CD bonds in the $\alpha/\omega$ positions, in close quantitative agreement with the excremental results of ref [24].

**Table III**. Calculated quadrupolar splittings in kHz, using $v_{q;CD} = 165\,\text{kHz}$.

|    | C1     | C2     | C3     | C4     |
|----|--------|--------|--------|--------|
| D  | -45.79 | -33.66 | -42.57 | -38.61 |
| D´ | -59.40 | -41.09 | -46.53 | -38.61 |

It is evident from the results in tables II and III that enantiotopic discrimination in the Nx phase of CB7CB is clearly accounted for by the simulations and, despite possible differences in the temperature, shows close quantitative agreement with experimental measurements, where available, and predicts the values of yet unmeasured spectral quantities. Moreover, the simulated segmental order parameters can be used to guide the assignment of spectral lines in certain cases of ambiguity. Notably, the quantitative agreement with experimental measurements on such detailed sets of quantities, that are sensitive to the local ordering within the phase, constitutes strong evidence that the simulated low temperature nematic phase of CB7CB conveys rather sensitively the molecular structure of the actual Nx phase formed by these dimers.

In closing this section we note that the size of the simulated system is necessarily large, in order to provide a consistent description of the modulated ordering. However it is not large enough to provide insights into the formation of domains of opposite helical sense. Due to the strict achirality of the constituent dimer molecules, both senses of helical twisting are in principle thermodynamically equivalent and there is amble experimental evidence[57] that chirally unbiased samples of the Nx phase consist indeed of domains showing opposite helical twisting. However, the estimated sizes of such domains are prohibitively large to allow their reproduction in atomistic detail simulations. Samples of opposite twisting sense, and fluctuations between the two senses, are of course observed in our simulations on cooling into the low temperature nematic phase, but once a given twisting sense is stabilized within the simulation box it remains unchanged on further cooling.

## 5. Conclusions.

The simulations of the CB7CB dimer reproduce successfully a number of experimentally well-established features of the Nx phase, such as its positional disorder, the length of the helical pitch and the doubling of the spectral lines associated with prochiral molecular sites, thus suggesting that the molecular structure described by these simulations applies correctly to the Nx phase. Furthermore, the fact that all the essential structural features found in these simulations can be accurately reproduced by the theoretical $N_{PT}$ model outlined in section 2, supports the consistency of the description of the Nx phase as exhibiting locally strong polar ordering of the molecules along a direction **m** that is helically modulated, remaining perpendicular to a "helix axis" $\mathbf{n}_h$. The latter axis constitutes macroscopically (i.e. over length scales exceeding substantially the helical pitch) an axis of uniaxial apolar symmetry. On the other hand, the description within the usual twist-bend ($N_{TB}$) model, being based on the modulation of the locally uniaxial and apolar nematic director **n**, fails to account even qualitatively for the crucial features found in the simulations. In particular, it directly implies apolar local ordering of the molecules and a common value of the "tilt" angle for the principal axes of all molecular tensors.

Regarding the emerging nematic polymorphism, it appears that the long sought thermotropic nematic phases with polar and biaxial ordering do materialize in the Nx phase of the symmetric achiral CBnCB dimers with odd-n spacer. However, unlike the envisaged ideal uniform polar and biaxial ordering, both the local polarity and the biaxiality appear as tightly modulated into structurally chiral domains within the Nx phase.

**Acknowledgements.**

Both authors had the pleasure of collaborating with Claudio Zannoni in a number of EU funded research projects over the last three decades and wish to express their appreciation for the friendship and kindness they have enjoyed in all their interactions with Claudio and the members of his research team in Bologna. The topic of the present invited paper was chosen in recognition of Claudio's immense contribution to the field of computer simulations of LCs.